\documentclass[proceedings]{JHEP37}
\usepackage{eurosym}
\usepackage{amssymb}
\usepackage{amsfonts}
\usepackage{amsmath}
\usepackage{epsfig}

\newbox\mybox

\newcommand\fverb{\setbox\mybox=\hbox\bgroup\verb}
\newcommand\fverbdo{\egroup\medskip\noindent\fbox{\unhbox\mybox}\ }
\newcommand\fverbit{\egroup\item[\fbox{\unhbox\mybox}]}
\conference{Metric versus observable operator representation}
\abstract{We elaborate further on the metric representation that is obtained by transferring the time-dependence from a Hermitian Hamiltonian to the metric operator in a related non-Hermitian
system. We provide further insight into the procedure on how to employ the time-dependent Dyson relation and the quasi-Hermiticity relation
to solve time-dependent Hermitian Hamiltonian systems. By solving both equations separately we argue here that it is in general easier to solve the former. 
We solve the mutually related time-dependent Schr{\"{o}}dinger equation for a Hermitian and non-Hermitian spin 1/2, 1 and 3/2 model with time-independent and time-dependent metric, 
respectively. In all models the overdetermined coupled system of equations for the Dyson map can be decoupled algebraic manipulations and reduces to simple linear differential equations and 
an equation that can be converted into the nonlinear Ermakov-Pinney equation.}

\title{Metric versus observable operator representation, higher spin models}
\author{Andreas Fring and Thomas Frith \\
Department of Mathematics, City, University of London,\\
Northampton Square, London EC1V 0HB, UK\\
E-mail: a.fring@city.ac.uk, thomas.frith@city.ac.uk}

\input{tcilatex}
\begin{document}

\section{Introduction}

Standard quantum mechanics allows for many equivalent variants to describe
the same physical observables. The well-known reason for this is that
expectation values are computed from ambiguous quantities in which the
individual components can be modified while the overall expression for the
expectation values are left unchanged. Gauge transformations are prominent
examples for such possible alterations. For time-dependent situations the
well known equivalence between the Schr\"{o}dinger and the Heisenberg
picture allows to change from time-dependent states and time-independent
operators to time-independent states and time-dependent operators,
respectively. Recently we \cite{fring2016exact} argued that in
time-dependent $\mathcal{PT}$-symmetric/quasi-Hermitian systems \cite%
{Bender:1998ke,Benderrev,Alirev} another variant is possible in which the
time-dependence is transferred from observables to metric operators. We will
refer to the former as the \emph{observable operator representation} and the
latter as the \emph{metric representation} by indicating the time-dependent
object in the name of the representation. These physically equivalent
representations are made possible in this setting as it always involves
non-trivial metric operators on the non-Hermitian side.

In \cite{fring2016exact} we demonstrated that the time-dependent Schr\"{o}%
dinger equation (TDSE) for a time-dependent Hermitian Hamiltonian, $%
h(t)=h^{\dagger }(t)$, and the easier TDSE for a time-independent Hermitian
Hamiltonian, $H\neq H^{\dagger }$, 
\begin{equation}
h(t)\phi (t)=i\hbar \partial _{t}\phi (t),\qquad \text{and\qquad }H\Psi
(t)=i\hbar \partial _{t}\Psi (t)  \label{TS}
\end{equation}%
may be treated equivalently. In the proposed scenario the Hermitian system
is governed by a time-dependent Hamiltonian $h(t)$ and a standard
time-independent metric operator $\mathbb{I}$, i.e. the unit operator,
whereas the non-Hermitian system is characterized by the time-independent
Hamiltonian $H$ and a non-standard time-dependent metric operator $\rho (t)$%
. The associated inner products in both systems are equivalent in the sense
that 
\begin{equation}
\left\langle \phi (t)\right. \left\vert \mathbb{I}\phi (t)\right\rangle
=\left\langle \Psi (t)\right. \left\vert \rho (t)\Psi (t)\right\rangle ,
\label{expt}
\end{equation}%
where the two wave functions $\phi (t)$ and $\Psi (t)$, solving the
respective equation in (\ref{TS}), are connected by the time-dependent
invertible Dyson operator $\eta (t)$ as%
\begin{equation}
\phi (t)=\eta (t)\Psi (t).  \label{sol}
\end{equation}%
The metric operator in (\ref{expt}) and the Dyson operator in (\ref{sol})
are simply related as $\rho (t):=\eta ^{\dagger }(t)\eta (t)$. Thus in this
picture the time-dependence has been moved from the Hamiltonian in the
Hermitian system to the metric operator in the non-Hermitian system.

There are two central equations that serve to determine the quantities
involved in the equations above. The first one, the time-dependent
quasi-Hermiticity relation%
\begin{equation}
H^{\dagger }\rho (t)-\rho (t)H=i\hbar \partial _{t}\rho (t),  \label{etaH}
\end{equation}%
results by demanding that the time-evolution is unitary, that is the
expectation values in (\ref{expt}) are preserved in time. Setting the time
derivative of (\ref{expt}) to zero and using the TDSE (\ref{TS}) leads to (%
\ref{etaH}). The second equations, the time-dependent Dyson relation 
\begin{equation}
h(t)=\eta (t)H\eta ^{-1}(t)+i\hbar \partial _{t}\eta (t)\eta ^{-1}(t),
\label{hH}
\end{equation}%
is obtained by substituting (\ref{sol}) into (\ref{TS}).

It was noted some time ago \cite{CA,CArev,time1,time6,fringmoussa} that as a
consequence of the Dyson relation (\ref{hH}) the Hamiltonian satisfying the
TDSE (\ref{TS}) is not observable\footnote{%
This fact can not be changed\ by imposing the additional constraint $i\hbar
\partial _{t}\eta (t)=\eta (t)H(t)$, as suggested in \cite{tolice}, as this
evidently produces a factor $2$. Moreover, this constraint implies that the
metric has to be time-independent so that $\eta (t)$ must be either
non-Hermitian or also time-independent.}, since observables $\mathcal{O}$ in
the non-Hermitian system need to be quasi-Hermitian, meaning they have to be
related to a corresponding observable $o$, i.e. a self-adjoint operator, in
the Hermitian system as $o(t)=\eta (t)\mathcal{O}(t)\eta ^{-1}(t)$. The
non-observability is also a feature when the Hamiltonian is explicitly
time-dependent, i.e. even for $H\rightarrow H(t)$. Furthermore, this implies
that $H$ is not the operator that characterizes the energy but instead the
operator 
\begin{equation}
\tilde{H}(t)=\eta ^{-1}(t)h(t)\eta (t)=H+i\hbar \eta ^{-1}(t)\partial
_{t}\eta (t),
\end{equation}%
that does not satisfy the TDSE, and is therefore by definition not a
Hamiltonian, is the energy operator in the non-Hermitian system. The
relation between the expectation values in the different systems is easily
verified to be 
\begin{equation}
\left\langle \phi (t)\right. \left\vert h(t)\phi (t)\right\rangle
=\left\langle \Psi (t)\left\vert \rho (t)\tilde{H}(t)\Psi (t)\right\rangle
\right. ,  \label{energy}
\end{equation}%
supporting the above statement. As demonstrated in \cite{fring2016exact}
unitary time-evolution operators $u(t,t^{\prime })$ and $U(t,t^{\prime })$
that evolves a state as $\phi (t)=u(t,t^{\prime })\phi (t^{\prime })$ or $%
\Psi (t)=U(t,t^{\prime })\Psi (t^{\prime })$, respectively, from a time $%
t^{\prime }$ to $t$ may also be constructed when $\phi (t)$ and $\Psi (t)$
have been obtained.

Since the equations (\ref{etaH}) and (\ref{hH}) describe highly
overdetermined systems it is a priori not evident whether they possess any
solutions at all and if they do whether they are meaningful. Remarkably such
solutions do exist and can be found as was demonstrated for time-dependent 
\cite{fringmoussa,fringmoussa2} and time-independent Hamiltonians \cite%
{fring2016exact}. Here we provide further solutions, focussing on the
limitations and in particular on the different solution procedures. In \cite%
{fring2016exact} we pursued the following process: Starting from a given a
non-Hermitian Hamiltonian $H$ we solved the time-dependent quasi-Hermiticity
relation (\ref{hH}) first, which seems most natural as it only involves one
unknown quantity, namely $\rho (t)$. Assuming the Dyson operator $\eta (t)$
to be Hermitian, it can in principle be computed from $\rho (t)$ by taking
its square root. Subsequently one may compute the Hermitian counterpart $%
h(t) $ by direct evaluation of the right hand side of\ the time-dependent
Dyson relation (\ref{hH}). As we will demonstrate in more detail below,
taking the square root in this case can be rather awkward and to avoid this
step we pursue here a different approach by solving the time-dependent Dyson
relation first. As we will see, this is more efficient, but evidently
requires some initial guess about the structure of the Hermitian Hamiltonian.

The models we consider here are slightly modified versions of the lattice
Yang-Lee model \cite{gehlen1,chainOla} 
\begin{equation}
H_{N}^{s}=-\frac{1}{2}\sum_{j=1}^{N}(c_{y}S_{j}^{y}+\omega c_{\omega }\vec{S}%
_{j}\cdot \vec{S}_{j+1}+ic_{x}\gamma S_{j}^{x}),\quad \omega ,\gamma
,c_{x},c_{y},c_{\omega }\in \mathbb{R},  \label{H}
\end{equation}%
where we allow for higher spin representations for the matrices $S_{j}^{x}$, 
$S_{j}^{y}$, $S_{j}^{z}$ at cite $j$ labelled by $s$. Our model parameters
are $\omega ,\gamma \in \mathbb{R}$ and the constants $c_{x},c_{y},c_{\omega
}$ are conveniently adjusted for the particular representations. Here we
will consider the one-site models and attempt, in analogy to the study in 
\cite{fring2016exact}, to map the non-Hermitian Hamiltonians to Hermitian
Hamiltonians of the form%
\begin{equation}
h(t)=-\frac{1}{2}\left[ \omega \mathbb{I}+\chi (t)S_{z}\right] ,  \label{ht}
\end{equation}%
where initially $\chi (t)$ is an arbitrary unknown function of $t$. It turns
out that in all spin models considered the time-dependent function $\chi (t)$
is restricted to obey an equation that can be converted easily into the
nonlinear Ermakov-Pinney equation.

\section{A solvable equivalence pair of spin 1/2 models}

The simplest version of $H_{N}^{s}$ is the one-site spin 1/2 model. Taking
the matrices $S_{j}^{x}$, $S_{j}^{y}$, $S_{j}^{z}$ simply to be the standard
Pauli spin matrices $\sigma _{j}^{x}$, $\sigma _{j}^{y}$, $\sigma _{j}^{z}$
and adjusting the constants $c_{x}=c_{y}=1$, $c_{\omega }=1/3$, the
Hamiltonian (\ref{H}) acquires the form%
\begin{equation}
H_{1}^{1/2}=-\frac{1}{2}(\sigma _{y}+\frac{\omega }{3}~\vec{\sigma}\cdot 
\vec{\sigma}+i\gamma \sigma _{x})=-\frac{1}{2}\left( 
\begin{array}{cc}
\omega & i(\gamma -1) \\ 
i(\gamma +1) & \omega%
\end{array}%
\right) .
\end{equation}%
The corresponding TDSE (\ref{TS}) is easily solved by 
\begin{equation}
\Psi _{\pm }(t)=\left( 
\begin{array}{c}
\pm i(1-\gamma ) \\ 
\phi%
\end{array}%
\right) e^{-itE_{\pm }},~~~\qquad E_{\pm }=-\frac{\omega }{2}\pm \frac{\phi 
}{2},
\end{equation}%
where $\phi :=\sqrt{1-\gamma ^{2}}$. Thus, this model exhibits the typical
feature for $\mathcal{PT}$-symmetric/quasi-Hermitian systems \cite%
{Benderrev,Alirev} that despite being described by a non-Hermitian
Hamiltonian there exists a range for the model parameters, in this case $%
\left\vert \gamma \right\vert \leq 1$, for which the eigenvalue spectrum is
real. Next we will solve the time-dependent Dyson relation\ (\ref{hH}) and
the time-dependent quasi-Hermiticity relation (\ref{etaH}) in more detail
and compare the advantages of one approach over the other.

\subsection{Solutions of the time-dependent quasi Hermiticity relation}

Assuming the time-dependent metric operator to be Hermitian we take it to be
in the most generic form 
\begin{equation}
\rho (t)=\left( 
\begin{array}{cc}
\rho _{1}(t) & \rho _{2}(t)-i\rho _{3}(t) \\ 
\rho _{2}(t)+i\rho _{3}(t) & \rho _{4}(t)%
\end{array}%
\right) ,
\end{equation}%
with unknown real functions $\rho _{i}$, $i=1,\ldots ,4$. Substituting this
Ansatz into (\ref{etaH}) and reading off the real and imaginary parts in
each matrix entry yields the four constraining first order differential
equations%
\begin{equation}
\dot{\rho}_{1}=(1+\gamma )\rho _{2},\qquad \dot{\rho}_{2}=\rho _{1}\frac{%
\gamma -1}{2}+\rho _{4}\frac{\gamma +1}{2},\qquad \dot{\rho}_{3}=0,\qquad 
\dot{\rho}_{4}=(\gamma -1)\rho _{2}.\qquad
\end{equation}%
As common we adopt the convention to indicate derivatives with respect to
time by an overdot. The general solution to these equations is easily
obtained as 
\begin{equation}
\rho _{1}(t)=\frac{1+\gamma }{\phi }\Gamma _{b_{2}}^{-b_{1}}+b_{4},\quad
\rho _{2}(t)=\Gamma _{b_{1}}^{b_{2}},\quad \rho _{3}(t)=b_{3},\quad \rho
_{4}(t)=\frac{1-\gamma }{\phi }\Gamma _{-b_{2}}^{b_{1}}+\frac{1-\gamma }{%
1+\gamma }b_{4},  \label{rho}
\end{equation}%
where we abbreviate $\Gamma _{x}^{y}:=x\sin (\phi t)+y\cos (\phi t)$ and
introduced the real integration constants $b_{1}$, $b_{2}$, $b_{3}$, $b_{4}$%
. To find (\ref{rho}) we just need to solve a harmonic oscillator equation
obtained from computing $\ddot{\rho}_{2}$ and the subsequent use the
expressions for $\dot{\rho}_{1}$, $\dot{\rho}_{4}$. Once $\rho _{2}$ is
known the remaining integrals are simply of first order.

In principle, we can take now the square root by diagonalizing $\rho $ first
as $\rho =UDU^{-1}$ and subsequently computing $\sqrt{\rho }=UD^{1/2}U^{-1}$%
. This is indeed feasible as shown in \cite{fring2016exact}, but even for
simple $2\times 2$-matrices it involves relatively lengthy expressions and
requires specific choices for the constants in order to guarantee that the
eigenvalues are all real. This is also the case for the model considered
here as seen from the determinant of $\rho $ 
\begin{equation}
\det \rho =\frac{1-\gamma }{1+\gamma }%
b_{4}^{2}-b_{1}^{2}-b_{2}^{2}-b_{3}^{2}.
\end{equation}%
Evidently this expression might become negative, so that at least one of the
two eigenvalues of $\rho $ would be negative and even for choices for which $%
\det \rho >0$ we may have two negative eigenvalues. We will not carry out
this step here, but instead follow an easier way to find $\eta $ from (\ref%
{hH}) and compare thereafter with the solution (\ref{rho}).

\subsection{Solutions of the time-dependent Dyson relation}

In order to solve the time-dependent Dyson relation we need to make some
pre-assumptions about the Hermitian Hamiltonian $h(t)$ and the map $\eta (t)$%
. We take $h(t)$ to be of the form as specified in (\ref{ht}) with $%
S_{z}=\sigma _{z}$ and assume $\eta (t)$ to be of the most generic Hermitian
form 
\begin{equation}
\eta (t)=\left( 
\begin{array}{cc}
\eta _{1}(t) & \eta _{2}(t)-i\eta _{3}(t) \\ 
\eta _{2}(t)+i\eta _{3}(t) & \eta _{4}(t)%
\end{array}%
\right) ,  \label{heta}
\end{equation}%
Taking this Ansatz into (\ref{hH}) leads to seven different constraining
equations%
\begin{equation}
\begin{array}{l}
\dot{\eta}_{1}=\eta _{2}\frac{\gamma +1}{2},~~~~~\dot{\eta}_{2}=\eta _{1}%
\frac{\gamma -1}{2}+\eta _{3}\frac{\chi }{2}=\eta _{4}\frac{\gamma +1}{2}%
+\eta _{3}\frac{\chi }{2},~~~~~\dot{\eta}_{3}=\eta _{2}\frac{\chi }{2},~~~~~%
\dot{\eta}_{4}=\eta _{2}\frac{\gamma -1}{2},%
\begin{array}{c}
\\ 
\\ 
\end{array}
\\ 
\eta _{3}(\gamma +1)-\chi \eta _{1}=\eta _{3}(1-\gamma )+\chi \eta _{4}=0.%
\end{array}%
\end{equation}%
Even though this system of equations is overdetermined, it can be solved by%
\begin{equation}
\eta _{1}(t)=\frac{c_{1}(\gamma +1)}{\chi ^{1/2}(\gamma -1)},\quad \eta
_{2}(t)=\frac{c_{1}\dot{\chi}}{\chi ^{3/2}(\gamma -1)},\quad \eta _{3}(t)=%
\frac{c_{1}\chi ^{1/2}}{\gamma -1},\quad \eta _{4}(t)=\frac{c_{1}}{\chi
^{1/2}},  \label{eta}
\end{equation}%
with one integration constant $c_{1}\in \mathbb{R}$ provided that the
function $\chi $ satisfies the second order nonlinear differential equation%
\begin{equation}
\ddot{\chi}-\frac{3}{2}\frac{\dot{\chi}^{2}}{\chi }-\frac{1}{2}\phi ^{2}\chi
+\frac{1}{2}\chi ^{3}=0.  \label{EPT}
\end{equation}%
Using the variable transformation $\chi =2/\sigma ^{2}$ this equation is
converted into the Ermakov-Pinney equation \cite{Ermakov,Pinney}%
\begin{equation}
\ddot{\sigma}+\frac{1}{4}\phi ^{2}\sigma =\frac{1}{\sigma ^{3}},  \label{EP}
\end{equation}%
which is ubiquitous in the context of the TDSE, e.g. \cite%
{Hone,Hawkins,ChoiK,ChoiK2,FBM,PhysRevD.90.084005}, and also some
quantization schemes \cite{Ioffe:2002tk,dey2015milne}. The general solution
to this equation is known to be%
\begin{equation}
\sigma (t)=\left[ A\sin ^{2}(\phi t/2)+B\cos ^{2}(\phi t/2)\pm 2C\sin (\phi
t/2)\cos (\phi t/2)\right] ^{1/2},
\end{equation}%
where the constants $A$, $B$ and $C$ are constraint as $AB-C^{2}=4/\phi ^{2}$%
, see \cite{eliezer}. Transforming back to $\chi $ and introducing the new
real constants $c_{2}$ and $c_{3}$ via the relations $A=2(-c_{3}\pm \sqrt{%
1+c_{2}^{2}+c_{3}^{2}})/\phi $ and $B=2(c_{3}\pm \sqrt{1+c_{2}^{2}+c_{3}^{2}}%
)/\phi $, we obtain the general solution to (\ref{EPT}) in the form%
\begin{equation}
\chi (t)=\frac{\phi }{c_{2}\sin (\phi t)+c_{3}\cos (\phi t)\pm \sqrt{%
1+c_{2}^{2}+c_{3}^{2}}}.
\end{equation}%
Thus with (\ref{eta}) and (\ref{heta}) we have obtained a generic solution
for $\eta $.

Let us now compare this with the solution of the time-dependent quasi
Hermiticity relation obtained in the previous subsection. Computing $\eta
^{2}$ from the above expressions and identifying the result as $\rho $ we
can compare with the solution (\ref{rho}) obtained previously. Matching the
constants as 
\begin{equation}
b_{1}=-\frac{2c_{3}\gamma c_{1}^{2}}{(1-\gamma )^{2}},~~b_{2}=\frac{%
2c_{2}\gamma c_{1}^{2}}{(1-\gamma )^{2}},~~b_{3}=\frac{2\gamma c_{1}^{2}}{%
(1-\gamma )^{2}},~~b_{4}=\frac{2\phi c_{1}^{2}\sqrt{1+c_{2}^{2}+c_{3}^{2}}}{%
(1-\gamma )^{3}},
\end{equation}%
the two solutions become identical. Evidently these constants could not have
been guessed in the approach of the previous subsection. With these values
the determinant becomes%
\begin{equation}
\det \rho =\frac{4(1+\gamma )c_{1}^{4}}{(1-\gamma )^{3}}%
(1+c_{2}^{2}+c_{3}^{2}),
\end{equation}%
which is positive for $\left\vert \gamma \right\vert \leq 1$. From the above
it is clear that it is far easier to solve (\ref{hH}) directly, as it can
essentially be reduced to some algebraic manipulations, a simple integration
and the Ermakov-Pinney equation for which the general solution is known.

We have now obtained all the ingredients to compute the solution to the TDSE
for the Hermitian system from (\ref{sol}). Assembling our results we obtain
from $\phi _{\pm }(t)=\sqrt{\mathcal{N}_{\pm }}\eta (t)\Psi _{\pm }(t)$ \
the normalized eigenvectors%
\begin{equation}
\phi _{\pm }(t)=c_{1}\sqrt{\mathcal{N}_{\pm }\chi }\left( 
\begin{array}{c}
\frac{1+\gamma }{\phi }i\left[ e^{\pm it(E_{+}-E_{-})}(ic_{2}\mp c_{3})+1\mp 
\sqrt{1+c_{2}^{2}+c_{3}^{2}}\right] \\ 
e^{\pm it(E_{+}-E_{-})}(\pm ic_{2}+c_{3})\pm 1-\sqrt{1+c_{2}^{2}+c_{3}^{2}}%
\end{array}%
\right) e^{-itE_{\pm }},
\end{equation}%
with normalization factors 
\begin{equation}
\mathcal{N}_{\pm }=\frac{1-\gamma }{\mp 4c_{1}^{2}\phi \left( \gamma \mp 
\sqrt{1+c_{2}^{2}+c_{3}^{2}}\right) }.
\end{equation}%
Form this we compute the expectation values%
\begin{eqnarray}
\mathcal{N}_{\pm }\left\langle \Psi _{\pm }(t)\right. \left\vert \rho
(t)\Psi _{\pm }(t)\right\rangle &=&\left\langle \phi _{\pm }(t)\right.
\left\vert \phi _{\pm }(t)\right\rangle =1, \\
\mathcal{N}_{\pm }\left\langle \Psi _{\pm }(t)\right. \left\vert \rho
(t)\Psi _{\mp }(t)\right\rangle &=&\left\langle \phi _{\pm }(t)\right.
\left\vert \phi _{\mp }(t)\right\rangle =\frac{\gamma \left( \pm
c_{3}+ic_{2}\right) {}}{\sqrt{\phi ^{2}+c_{2}^{2}+c_{3}^{2}}},
\end{eqnarray}%
which confirm that the time-evolution is indeed unitary. We also confirm the
validity of the relation for the energy expectations (\ref{energy}) by
computing%
\begin{equation}
\left\langle \phi _{\pm }(t)\right. \left\vert h(t)\phi _{\pm
}(t)\right\rangle =\mathcal{N}_{\pm }\left\langle \Psi _{\pm }(t)\left\vert
\rho (t)\tilde{H}(t)\Psi _{\pm }(t)\right\rangle \right. =\frac{\pm \phi ^{2}%
\sqrt{1+c_{2}^{2}+c_{3}^{2}}-c_{2}^{2}-c_{3}^{2}}{2(\phi
^{2}+c_{2}^{2}+c_{3}^{2})}\chi (t)-\frac{\omega }{2}.
\end{equation}

While we found some explicit solutions, this example also demonstrates that
one can not map to any arbitrary given target Hamiltonian, as $\chi (t)$ is
restricted by the nonlinear equation (\ref{EPT}).

\section{A solvable equivalence pair of spin 1 models}

Increasing the dimension of the spin representation poses a more difficult
challenge, but as we will see many of the features we observed for the spin
1/2 model will survive. Let us next consider a generalization of the
previous model to a spin 1 model where the matrices $S_{j}^{x}$, $S_{j}^{y}$%
, $S_{j}^{z}$ in (\ref{H}) are taken to be the standard $3\times 3$-spin 1
matrices 
\begin{equation}
S^{x}=\frac{1}{\sqrt{2}}\left( 
\begin{array}{ccc}
0 & 1 & 0 \\ 
1 & 0 & 1 \\ 
0 & 1 & 0%
\end{array}%
\right) ,\quad \quad S^{y}=\frac{1}{\sqrt{2}}\left( 
\begin{array}{ccc}
0 & -i & 0 \\ 
i & 0 & -i \\ 
0 & i & 0%
\end{array}%
\right) ,\quad \quad S^{z}=\left( 
\begin{array}{ccc}
1 & 0 & 0 \\ 
0 & 0 & 0 \\ 
0 & 0 & -1%
\end{array}%
\right) .
\end{equation}%
Choosing the constants $c_{x},c_{y},c_{\omega }$ conveniently this
Hamiltonian simplifies for $N=1$ to 
\begin{equation}
H_{1}^{1}=-\frac{1}{\sqrt{2}}(S^{y}+\frac{\omega }{\sqrt{2}}\mathbb{I}%
+i\gamma S^{x})=-\frac{1}{2}\left( 
\begin{array}{ccc}
\omega & i(\gamma -1) & 0 \\ 
i(\gamma +1) & \omega & i(\gamma -1) \\ 
0 & i(\gamma +1) & \omega%
\end{array}%
\right) .  \label{H1}
\end{equation}%
The corresponding TDSE (\ref{TS}) is solved by 
\begin{equation}
\Psi _{k}(t)=\left( 
\begin{array}{c}
(-1)^{k}(1-\gamma ) \\ 
2ik\tilde{\phi} \\ 
1-\gamma%
\end{array}%
\right) e^{-itE_{k}},~~~\qquad E_{k}=-\frac{\omega }{2}+k\tilde{\phi},\quad
~~~k=0,\pm 1
\end{equation}%
where $\tilde{\phi}:=\sqrt{(1-\gamma ^{2})/2}$. Once again in the parameter
region $\left\vert \gamma \right\vert \leq 1$ the non-Hermitian Hamiltonian (%
\ref{H1}) possesses a real eigenvalue spectrum. Next we solve (\ref{etaH})
and (\ref{hH}).

\subsection{Solutions of the time-dependent quasi Hermiticity relation}

Assuming the time-dependent metric operator to be Hermitian we substitute
the most generic Ansatz 
\begin{equation}
\rho (t)=\left( 
\begin{array}{ccc}
\rho _{1}(t) & \rho _{2}(t)-i\rho _{3}(t) & \rho _{4}(t)-i\rho _{5}(t) \\ 
\rho _{2}(t)+i\rho _{3}(t) & \rho _{6}(t) & \rho _{7}(t)-i\rho _{8}(t) \\ 
\rho _{4}(t)+i\rho _{5}(t) & \rho _{7}(t)+i\rho _{8}(t) & \rho _{9}(t)%
\end{array}%
\right) .
\end{equation}%
into the time-dependent quasi Hermiticity relation (\ref{etaH}) obtaining in
principle $18$ equation for the nine real functions $\rho _{i}(t),$ $%
i=1,\ldots ,9$. Excluding vanishing and related ones we are left with nine
equations%
\begin{equation}
\begin{array}{l}
\dot{\rho}_{1}=\rho _{2}(\gamma +1),\quad \dot{\rho}_{2}=\rho _{1}\frac{%
\gamma -1}{2}+(\rho _{4}+\rho _{6})\frac{\gamma +1}{2},\quad \dot{\rho}%
_{3}=\rho _{5}\frac{\gamma +1}{2}, \\ 
\dot{\rho}_{4}=\rho _{2}\frac{\gamma -1}{2}+\rho _{7}\frac{\gamma +1}{2}%
,\quad \dot{\rho}_{5}=\rho \frac{\gamma -1}{2}+\rho _{8}\frac{\gamma +1}{2}%
,\quad \dot{\rho}_{6}=\rho _{2}(\gamma -1)+\rho _{7}(\gamma +1)%
\begin{array}{c}
\\ 
\\ 
\end{array}
\\ 
\dot{\rho}_{7}=(\rho _{4}+\rho _{6})\frac{\gamma -1}{2}+\rho _{9}\frac{%
\gamma +1}{2},\quad \dot{\rho}_{8}=\rho _{5}\frac{\gamma -1}{2},\quad \dot{%
\rho}_{9}=\rho _{7}(\gamma -1).%
\end{array}%
\end{equation}%
Once again as in the spin $1/2$ case we have as many equations as unknown
functions and it is straightforward to solve these equations, as
substitutions lead to simple integrals. We find the solutions 
\begin{equation}
\begin{array}{l}
\rho _{1}(t)=\frac{\left( 2b_{4}+3b_{5}\right) (\gamma +1)}{8(1-\gamma )}+%
\tilde{\Gamma}_{b_{6}}^{b_{7}}+\breve{\Gamma}_{b_{8}}^{b_{9}},\quad \quad
\rho _{2}(t)=\frac{\phi }{1+\gamma }\left[ \tilde{\Gamma}_{-b_{7}}^{b_{6}}+2%
\breve{\Gamma}_{-b_{9}}^{b_{8}}\right] ,\quad \\ 
\rho _{3}(t)=\frac{1+\gamma }{2\phi }\left[ \tilde{\Gamma}_{b_{2}}^{-b_{1}}%
\right] +b_{3},\quad \quad \rho _{4}(t)=\frac{\gamma -1}{\gamma +1}\breve{%
\Gamma}_{b_{8}}^{b_{9}}+\frac{1}{8}(6b_{4}+b_{5}),\quad \quad \rho _{5}(t)=%
\tilde{\Gamma}_{b_{1}}^{b_{2}}, \\ 
\rho _{6}(t)=2\frac{\gamma -1}{\gamma +1}\breve{\Gamma}_{b_{8}}^{b_{9}}-%
\frac{1}{4}(2b_{4}-b_{5}),\quad \quad \rho _{7}(t)=\frac{1}{\sqrt{2}}\left( 
\frac{1-\gamma }{1+\gamma }\right) ^{3/2}\left[ \tilde{\Gamma}%
_{-b_{7}}^{b_{6}}+2\breve{\Gamma}_{b_{9}}^{-b_{8}}\right] , \\ 
\rho _{8}(t)=\frac{\phi }{1+\gamma }\tilde{\Gamma}_{-b_{2}}^{b_{1}}+\frac{%
1-\gamma }{1+\gamma }b_{3},\quad \quad \rho _{9}(t)=\left( \frac{1-\gamma }{%
1+\gamma }\right) ^{2}\left[ -\tilde{\Gamma}_{b_{6}}^{b_{7}}+\breve{\Gamma}%
_{b_{8}}^{b_{9}}\right] +\frac{1-\gamma }{8(1+\gamma )}(2b_{4}+3b_{5}),%
\end{array}%
\end{equation}%
with nine integration constants $b_{i},$ $i=1,\ldots ,9$. We abbreviated $%
\tilde{\Gamma}_{x}^{y}:=x\sin (\tilde{\phi}t)+y\cos (\tilde{\phi}t)$ and $%
\breve{\Gamma}_{x}^{y}:=x\sin (2\tilde{\phi}t)+y\cos (2\tilde{\phi}t)$. For
this solution it is even less evident to chose suitable constants and
simplifying choices by setting some of the $b_{i}$ to zero usually yield
negative eigenvalues for $\rho $. Thus we will not compute the root, but
return to this solution below for comparison.

\subsection{Solutions of the time-dependent Dyson relation}

Instead we solve the time-dependent Dyson equation (\ref{hH}). We assume a
similar form for our Hermitian target Hamiltonian as in (\ref{ht}) and take $%
S^{z}$ to be a spin 1 matrix, denote $\chi =X$ and take $\eta (t)$ to be of
the Hermitian form 
\begin{equation}
\eta (t)=\left( 
\begin{array}{ccc}
\eta _{1}(t) & \eta _{2}(t)-i\eta _{3}(t) & \eta _{4}(t)-i\eta _{5}(t) \\ 
\eta _{2}(t)+i\eta _{3}(t) & \eta _{6}(t) & \eta _{7}(t)-i\eta _{8}(t) \\ 
\eta _{4}(t)+i\eta _{5}(t) & \eta _{7}(t)+i\eta _{8}(t) & \eta _{9}(t)%
\end{array}%
\right) .
\end{equation}%
Substituting these expressions into the time-dependent Dyson equation (\ref%
{hH}) yields in principle $18$ equation for the real functions $\eta
_{i}(t), $ $i=1,\ldots ,9$. We obtain%
\begin{equation}
\begin{array}{l}
\dot{\eta}_{1}=\eta _{2}\frac{\gamma +1}{2},\quad ~\dot{\eta}_{2}=\eta _{1}%
\frac{\gamma -1}{2}+\eta _{3}\frac{X}{2}+\eta _{4}\frac{\gamma +1}{2}=\eta
_{6}\frac{\gamma +1}{2},\quad ~\dot{\eta}_{3}=-\eta _{2}\frac{X}{2}+\eta _{5}%
\frac{\gamma +1}{2}=0, \\ 
\dot{\eta}_{4}=\eta _{2}\frac{\gamma -1}{2}+\eta _{5}\frac{X}{2}=\eta _{5}%
\frac{X}{2}+\eta _{7}\frac{\gamma +1}{2},\quad \quad \dot{\eta}_{5}=\eta _{3}%
\frac{\gamma -1}{2}-\eta _{4}\frac{X}{2}=\eta _{4}\frac{X}{2}-\eta _{8}\frac{%
\gamma +1}{2},%
\begin{array}{c}
\\ 
\\ 
\end{array}
\\ 
\dot{\eta}_{6}=\eta _{2}\frac{\gamma -1}{2}+\eta _{7}\frac{\gamma +1}{2}%
,\quad \quad \dot{\eta}_{7}=\eta _{4}\frac{\gamma -1}{2}+\eta _{8}\frac{X}{2}%
+\eta _{9}\frac{\gamma +1}{2}=\eta _{6}\frac{\gamma -1}{2}, \\ 
\dot{\eta}_{8}=\eta _{5}\frac{\gamma -1}{2}-\eta _{7}\frac{X}{2}=0,\quad
\quad \dot{\eta}_{9}=\eta _{7}\frac{\gamma -1}{2},%
\begin{array}{c}
\\ 
\\ 
\end{array}%
\end{array}%
\end{equation}%
and%
\begin{equation}
(1+\gamma )\eta _{3}-X\eta _{1}=(1-\gamma )\eta _{3}+(1+\gamma )\eta
_{8}=(1-\gamma )\eta _{8}+X\eta _{9}=0.  \label{eta10}
\end{equation}%
Unlike the system of equations for the metric operator this set is
overdetermined. Nonetheless, they may be solved by%
\begin{equation}
\begin{array}{l}
\eta _{1}(t)=\frac{c_{1}}{X},~\quad \eta _{2}(t)=-\frac{2c_{1}\dot{X}}{%
(1+\gamma )^{2}X^{2}},\quad ~\eta _{3}(t)=\frac{c_{1}}{1+\gamma },\quad
~\eta _{4}(t)=\frac{c_{1}(4\dot{X}^{2}-X^{4})}{2(1+\gamma )^{2}X^{3}}, \\ 
\eta _{5}(t)=-\frac{2c_{1}\dot{X}}{(1+\gamma )^{2}X},\quad ~\eta _{6}(t)=%
\frac{c_{1}(4\dot{X}^{2}+X^{4}-4\phi ^{2}X^{2})}{2(1+\gamma )^{2}X^{3}}%
,\quad ~\eta _{7}(t)=\frac{2(1-\gamma )c_{1}\dot{X}}{(1+\gamma )^{2}X^{2}}%
,\quad \quad 
\begin{array}{c}
\\ 
\\ 
\end{array}
\\ 
\eta _{8}(t)=\frac{c_{1}(\gamma -1)}{(\gamma +1)^{2}},\quad ~\eta _{9}(t)=%
\frac{c_{1}(1-\gamma )^{2}}{(1+\gamma )^{2}X},%
\end{array}%
\end{equation}%
where $X(t)$ is restricted to obey the second order non-linear differential
equation%
\begin{equation}
\ddot{X}-\frac{3}{2}\frac{\dot{X}^{2}}{X}-\frac{1}{2}\tilde{\phi}^{2}X+\frac{%
X^{3}}{8}=0.  \label{chid}
\end{equation}%
This equations closely resembles (\ref{EPT}) and we can once more transform
it to the Ermakov-Pinney equation (\ref{EP}) with $\sigma \rightarrow \tilde{%
\sigma}$, $\phi \rightarrow \tilde{\phi}$ by using $X=4/\tilde{\sigma}^{2}$
in this case. Following the same steps of the previous subsection we obtain
the general solution for (\ref{chid}) as 
\begin{equation}
X(t)=\frac{2\tilde{\phi}}{c_{2}\sin (\tilde{\phi}t)+c_{3}\cos (\tilde{\phi}%
t)\pm \sqrt{1+c_{2}^{2}+c_{3}^{2}}}.
\end{equation}%
Again we compute $\eta ^{2}$ and compare the result with $\rho $ from the
previous subsection. Identifying the constants as 
\begin{equation}
\begin{array}{l}
b_{1}=-\frac{4\gamma ^{2}c_{1}^{2}c_{3}}{(1+\gamma )^{4}},~~b_{2}=\frac{%
4\gamma ^{2}c_{1}^{2}c_{2}}{(1+\gamma )^{4}},~~b_{3}=\frac{2\gamma c_{1}^{2}%
\sqrt{1+c_{2}^{2}+c_{3}^{2}}}{\phi (1+\gamma )^{3}},\quad b_{4}=\frac{%
2c_{1}^{2}\left[ \phi ^{2}(3-c_{2}^{2}-c_{3}^{2})-2\right] }{(1+\gamma )^{4}}%
, \\ 
b_{5}=\frac{2c_{1}^{2}(3+\gamma ^{2})(1+c_{2}^{2}+c_{3}^{2})}{(1+\gamma )^{4}%
},\quad b_{6}=\frac{2\gamma c_{1}^{2}c_{2}\sqrt{1+c_{2}^{2}+c_{3}^{2}}}{\phi
^{2}(1+\gamma )^{2}},~~b_{7}=\frac{2\gamma c_{1}^{2}c_{3}\sqrt{%
1+c_{2}^{2}+c_{3}^{2}}}{\phi ^{2}(1+\gamma )^{2}},~~~~%
\begin{array}{c}
\\ 
\\ 
\end{array}
\\ 
b_{8}=\frac{\gamma ^{2}c_{1}^{2}c_{2}c_{3}}{\phi ^{2}(1+\gamma )^{2}},\quad
b_{9}=\frac{\gamma ^{2}c_{1}^{2}(c_{2}^{2}-c_{3}^{2})}{2\phi ^{2}(1+\gamma
)^{2}},%
\end{array}
\label{b8}
\end{equation}%
the two solutions coincide. This demonstrates once more why simple choices
for the constants $c_{i}$ did not yield meaningful solutions for $\eta $.
Using the values (\ref{b8}) we compute the determinant%
\begin{equation}
\det \rho =8\frac{(1-\gamma )^{3}}{(1+\gamma )^{9}}%
c_{1}^{6}(1+c_{2}^{2}+c_{3}^{2})^{3},
\end{equation}%
which is positive for the parameter range of interest.

Having computed the Dyson map and the solution to the TDSE for $H$ we obtain
the solution for the TDSE involving $h(t)$ from (\ref{sol})%
\begin{equation}
\begin{array}{l}
\phi _{\pm }(t)=\left( 
\begin{array}{c}
-X(t)\left[ 1\mp \sqrt{1+c_{2}^{2}+c_{3}^{2}}+e^{it(E_{\pm
}-E_{0})}(ic_{2}\mp c_{3})\left[ 1\mp 2\phi /X(t)\right] \right] \\ 
2(1-\gamma )(-c_{2}\mp ic_{3})e^{it(E_{\pm }-E_{0})}%
\begin{array}{c}
\\ 
\\ 
\end{array}
\\ 
(1-\gamma )X(t)\left[ 1\pm \sqrt{1+c_{2}^{2}+c_{3}^{2}}-e^{it(E_{\pm
}-E_{0})}(ic_{2}\mp c_{3})\left[ 1\pm 2\phi /X(t)\right] \right]%
\end{array}%
\right) \frac{c_{1}e^{-itE_{\pm }}}{1+\gamma }%
\begin{array}{c}
\\ 
\\ 
\\ 
\\ 
\\ 
\\ 
\end{array}
\\ 
\phi _{0}(t)=\left( 
\begin{array}{c}
X(t)\left[ ic_{3}\sin (\phi t)-ic_{2}\cos (\phi t)-1\right] +2\phi \sqrt{%
1+c_{2}^{2}+c_{3}^{2}} \\ 
2i(1-\gamma )%
\begin{array}{c}
\\ 
\\ 
\end{array}
\\ 
(1-\gamma )\left[ X(t)\left[ -ic_{3}\sin (\phi t)+ic_{2}\cos (\phi t)-1%
\right] +2\phi \sqrt{1+c_{2}^{2}+c_{3}^{2}}\right]%
\end{array}%
\right) \frac{c_{1}e^{-itE_{0}}}{1+\gamma }%
\end{array}%
\end{equation}
As in the previous section we can use these expressions to confirm that the
time-evolution is unitary and also verify (\ref{energy}) .

\section{A solvable equivalence pair of spin 3/2 models}

Finally we also consider a spin 3/2 model and take the matrices $S_{j}^{x}$, 
$S_{j}^{y}$, $S_{j}^{z}$ in (\ref{H}) to be $4\times 4$-spin 3/2 matrices 
\begin{equation}
S^{x}=\frac{1}{2}\left( 
\begin{array}{cccc}
0 & \sqrt{3} & 0 & 0 \\ 
\sqrt{3} & 0 & 2 & 0 \\ 
0 & 2 & 0 & \sqrt{3} \\ 
0 & 0 & \sqrt{3} & 0%
\end{array}%
\right) ,~S^{y}=\frac{i}{2}\left( 
\begin{array}{cccc}
0 & -\sqrt{3} & 0 & 0 \\ 
\sqrt{3} & 0 & -2 & 0 \\ 
0 & 2 & 0 & -\sqrt{3} \\ 
0 & 0 & \sqrt{3} & 0%
\end{array}%
\right) ,~S^{z}=\frac{1}{2}\left( 
\begin{array}{cccc}
3 & 0 & 0 & 0 \\ 
0 & 1 & 0 & 0 \\ 
0 & 0 & -1 & 0 \\ 
0 & 0 & 0 & -3%
\end{array}%
\right) ,
\end{equation}%
at cite $j$. Choosing the constants $c_{x},c_{y},c_{\omega }$ conveniently,
for $N=1$ this Hamiltonian simplifies to 
\begin{equation}
H_{1}^{3/2}=-\frac{1}{6}(S^{y}+\frac{2\omega }{3}\mathbb{I}+i\gamma S^{x})=-%
\frac{1}{4}\left( 
\begin{array}{cccc}
\omega & i\frac{\gamma -1}{\sqrt{3}} & 0 & 0 \\ 
i\frac{\gamma -1}{\sqrt{3}} & \omega & i\frac{\gamma -1}{6} & 0 \\ 
0 & i\frac{\gamma -1}{6} & \omega & \frac{\gamma -1}{\sqrt{3}} \\ 
0 & 0 & \frac{\gamma -1}{\sqrt{3}} & \omega%
\end{array}%
\right) .
\end{equation}%
The corresponding TDSE (\ref{TS}) is solved to 
\begin{equation}
\Psi _{k}(t)=\left( 
\begin{array}{c}
i(1-\gamma )^{3/2} \\ 
-2\sqrt{3}k\hat{\phi}(1-\gamma )^{1/2} \\ 
2i\sqrt{3}(k^{2}-2\left\vert k\right\vert )\hat{\phi}(1+\gamma )^{1/2} \\ 
\limfunc{sign}(k)(\left\vert k\right\vert -2)(1+\gamma )^{3/2}%
\end{array}%
\right) e^{-itE_{k}},~~~\qquad E_{k}=-\frac{1}{2}k~\hat{\phi}-\frac{\omega }{%
4},\quad ~~~k=\pm 1,\pm 3
\end{equation}%
where $\hat{\phi}:=\sqrt{1-\gamma ^{2}}/6$. The eigenvalue spectrum is real
for the same parameter range as in the previous subsections.

Here we will only solve the time-dependent Dyson equation (\ref{hH}) to see
whether the features of the spin 1/2 and spin 1 models are also present in
this model. We assume a similar form for our Hermitian target Hamiltonian as
in (\ref{ht}), denote $\chi =\Xi $ and take $\eta (t)$ to be of the
Hermitian form 
\begin{equation}
\eta (t)=\left( 
\begin{array}{cccc}
\eta _{1}(t) & \eta _{2}(t)-i\eta _{3}(t) & \eta _{4}(t)-i\eta _{5}(t) & 
\eta _{6}(t)-i\eta _{7}(t) \\ 
\eta _{2}(t)+i\eta _{3}(t) & \eta _{8}(t) & \eta _{9}(t)-i\eta _{10}(t) & 
\eta _{11}(t)-i\eta _{12}(t) \\ 
\eta _{4}(t)+i\eta _{5}(t) & \eta _{7}(t)+i\eta _{8}(t) & \eta _{13}(t) & 
\eta _{14}(t)-i\eta _{15}(t) \\ 
\eta _{6}(t)+i\eta _{7}(t) & \eta _{11}(t)+i\eta _{12}(t) & \eta
_{14}(t)+i\eta _{15}(t) & \eta _{16}(t)%
\end{array}%
\right) .
\end{equation}%
Substituting these expressions into the time-dependent Dyson equation (\ref%
{hH}) yields in principle $32$ equation for the $\eta _{i}(t),$ $i=1,\ldots
,16$. Once again the system is highly overdetermined, but remarkably it can
be solved similarly as in the previous sections. Here we only present the
solutions to these equations. We find%
\begin{equation}
\begin{array}{l}
\eta _{1}(t)=\frac{c_{1}}{\Xi ^{3/2}},~\quad \eta _{2}(t)=-\frac{6\sqrt{3}%
c_{1}\dot{X}}{(1+\gamma )\Xi ^{5/2}},\quad ~\eta _{3}(t)=\frac{3\sqrt{3}c_{1}%
}{(1+\gamma )\Xi ^{1/2}},\quad ~\eta _{4}(t)=\frac{9\sqrt{3}c_{1}(4\dot{\Xi}%
^{2}-\Xi ^{4})}{(1+\gamma )^{2}\Xi ^{7/2}}, \\ 
\eta _{5}(t)=-\frac{36\sqrt{3}c_{1}\dot{\Xi}}{(1+\gamma )^{2}\Xi ^{3/2}}%
,\quad ~\eta _{6}(t)=\frac{54c_{1}(3\dot{\Xi}\Xi ^{4}-4\dot{\Xi}^{3})}{%
(1+\gamma )^{3}\Xi ^{9/2}},\quad ~\eta _{7}(t)=\frac{27c_{1}(12\dot{\Xi}%
^{2}-\Xi ^{4})}{(1+\gamma )^{3/2}\Xi ^{5/2}},\quad \quad 
\begin{array}{c}
\\ 
\\ 
\end{array}
\\ 
\eta _{8}(t)=\frac{6c_{1}(12\dot{\Xi}^{2}+3\Xi ^{4}-\hat{\phi}^{2}\Xi ^{2})}{%
(\gamma +1)^{2}\Xi ^{7/2}},\quad ~\eta _{9}(t)=\frac{18c_{1}\dot{\Xi}(4\hat{%
\phi}^{2}\Xi ^{2}-12\dot{\Xi}^{2}-3\Xi ^{4})}{(\gamma +1)^{3}\Xi ^{9/2}}%
,\quad  \\ 
\eta _{10}(t)=\frac{9c_{1}\dot{\Xi}(4\hat{\phi}^{2}\Xi ^{2}-12\dot{\Xi}%
^{2}-3\Xi ^{4})}{(\gamma +1)^{3}\Xi ^{5/2}},\quad \eta _{11}(t)=\frac{9\sqrt{%
3}c_{1}(1-\gamma )(\Xi ^{4}-4\dot{X}^{2})}{(1+\gamma )^{3}\Xi ^{7/2}},\quad ~%
\begin{array}{c}
\\ 
\\ 
\end{array}
\\ 
\eta _{12}(t)=\frac{36\sqrt{3}c_{1}(1-\gamma )\dot{\Xi}}{(1+\gamma )^{3}\Xi
^{3/2}},\quad \eta _{13}(t)=\frac{6c_{1}(\gamma -1)(12\dot{\Xi}^{2}+3\Xi
^{4}-\hat{\phi}^{2}\Xi ^{2})}{(1+\gamma )^{3}\Xi ^{7/2}},\quad ~ \\ 
\eta _{14}(t)=-\frac{6\sqrt{3}c_{1}(1-\gamma )^{2}\dot{\Xi}}{(1+\gamma
)^{3}\Xi ^{5/2}},\quad ~\eta _{15}(t)=\frac{3\sqrt{3}c_{1}(1-\gamma )^{2}}{%
(1+\gamma )^{3}\Xi ^{1/2}},\quad \eta _{16}(t)=\frac{c_{1}(\gamma -1)3}{%
(1+\gamma )^{3}\Xi ^{3/2}},%
\begin{array}{c}
\\ 
\\ 
\end{array}%
\end{array}%
\end{equation}%
where $\Xi (t)$ has to obey the second order non-linear differential equation%
\begin{equation}
\ddot{\Xi}-\frac{3}{2}\frac{\dot{\Xi}^{2}}{\Xi }-\frac{1}{2}\hat{\phi}%
^{2}\Xi +\frac{\Xi ^{3}}{8}=0.
\end{equation}%
As in the previous subsection we can transform it to the Ermakov-Pinney
equation (\ref{EP}) with $\sigma \rightarrow \hat{\sigma}$, $\phi
\rightarrow \hat{\phi}$ using $\Xi =4/\hat{\sigma}^{2}$ in this case and
therefore we have 
\begin{equation}
\Xi (t)=\frac{2\hat{\phi}}{c_{2}\sin (\hat{\phi}t)+c_{3}\cos (\hat{\phi}%
t)\pm \sqrt{1+c_{2}^{2}+c_{3}^{2}}}.
\end{equation}%
Computing from this $\rho =\eta ^{2}$, we evaluate the determinant to%
\begin{equation}
\det \rho =\frac{6^{6}(1-\gamma )^{6}c_{1}^{8}}{(1+\gamma )^{18}}%
(1+c_{2}^{2}+c_{3}^{2})^{6},
\end{equation}%
which is always positive for the parameter range of interest. Naturally (\ref%
{sol}) yields once more the soltution to the TDSE for $h(t)$.

\section{Conclusions}

We have demonstrated that metric representations lead to consistent
descriptions equivalent to the operator representation by providing further
solutions to the time-dependent quasi Hermiticity relation (\ref{etaH}) and
the time-dependent Dyson relation (\ref{hH}). For the spin models we
considered here we observed that the determining relation for the metric
operator (\ref{etaH})\ converts into as many equations as unknown functions.
The equations are easily decoupled and integrated to determine the metric
operator. However, the diagonalization needed in order to take the square
root is usually and moreover requires specific choices for the constants
involved to ensure the all eigenvalues are positive. As we have demonstrated
simple choices are usually not evident or do not even exist. In order to
bypass this step we pursued what turned out to be an easier approach and
solved the time-dependent Dyson relation (\ref{hH}) instead. Assuming a
general form for the Hermitian Hamiltonian in (\ref{hH}) converts it into an
overdetermined set of equations for the components of the Dyson map.
Remarkably these equations can be decoupled and solved by simple
integrations for the components of $\eta $. The time-dependent equation
occurring in the Hermitian Hamiltonian is restricted by a nonlinear equation
that can be converted into the Ermakov-Pinney equation. This feature was
observed in all three spin models considered here and based on this
observation we conjecture that it might be universal and will hold for all
higher spin representations.

Evidently there are many interesting open problems left for future research,
such as a more extensive treatment of systems with explicitly time-dependent
non-Hermitian Hamiltonian and with regard to the spin models more sites pose
a natural challenge.

\bigskip \noindent \textbf{Acknowledgments:} TF is supported by a City,
University of London Research Fellowship.

\newif\ifabfull\abfulltrue


\begin{thebibliography}{99}
\bibitem{fring2016exact} A.~Fring and T.~Frith, \newblock Exact analytical
solutions for time-dependent Hermitian Hamiltonian  systems from static
unobservable non-Hermitian Hamiltonians, \newblock arXiv:1610.07537 (2016).

\bibitem{Bender:1998ke} C.~M. Bender and S.~Boettcher, \newblock Real
Spectra in Non-Hermitian Hamiltonians Having PT Symmetry, \newblock Phys.
Rev. Lett. \textbf{80}, 5243--5246 (1998).

\bibitem{Benderrev} C.~M. Bender, \newblock Making sense of non-Hermitian
Hamiltonians, \newblock Rept. Prog. Phys. \textbf{70}, 947--1018 (2007).

\bibitem{Alirev} A.~Mostafazadeh, \newblock Pseudo-Hermitian Representation
of Quantum Mechanics, \newblock Int. J. Geom. Meth. Mod. Phys. \textbf{7},
1191--1306 (2010).

\bibitem{CA} C.~Figueira~de Morisson~Faria and A.~Fring, \newblock Time
evolution of non-Hermitian Hamiltonian systems, \newblock J. Phys. \textbf{%
A39}, 9269--9289 (2006).

\bibitem{CArev} C.~Figueira~de Morisson~Faria and A.~Fring, \newblock %
Non-Hermitian Hamiltonians with real eigenvalues coupled to electric 
fields: from the time-independent to the time dependent quantum mechanical 
formulation, \newblock Laser Physics \textbf{17}, 424--437 (2007).

\bibitem{time1} A.~Mostafazadeh, \newblock Time-dependent pseudo-Hermitian
Hamiltonians defining a unitary  quantum system and uniqueness of the metric
operator, \newblock Physics Letters B \textbf{650}(2), 208--212 (2007).

\bibitem{time6} M.~Znojil, \newblock Time-dependent version of
crypto-Hermitian quantum theory, \newblock Physical Review D \textbf{78}(8),
085003 (2008).

\bibitem{fringmoussa} A.~Fring and M.~H.~Y. Moussa, \newblock Unitary
quantum evolution for time-dependent quasi-Hermitian systems  with
nonobservable Hamiltonians, \newblock Physical Review A \textbf{93}(4),
042114 (2016).

\bibitem{tolice} F.~S. Luiz, M.~A. Pontes, and M.~H.~Y. Moussa, \newblock %
Unitarity of the time-evolution and observability of non-Hermitian 
Hamiltonians for time-dependent Dyson maps, \newblock arXiv:1611.08286
(2016).

\bibitem{fringmoussa2} A.~Fring and M.~M. H.~Y. Moussa, \newblock %
Non-Hermitian Swanson model with a time-dependent metric, \newblock Physical
Review A \textbf{94}(4), 042128 (2016).

\bibitem{gehlen1} G.~von Gehlen, 
\newblock {Critical and off critical conformal analysis of the Ising quantum
  chain in an imaginary field}, \newblock J. Phys. \textbf{A24}, 5371--5400
(1991).

\bibitem{chainOla} O.~A. Castro-Alvaredo and A.~Fring, \newblock A spin
chain model with non-Hermitian interaction: The Ising quantum  spin chain in
an imaginary field, \newblock J. Phys. \textbf{A42}, 465211 (2009).

\bibitem{Ermakov} V.~Ermakov, \newblock Transformation of differential
equations,, \newblock Univ. Izv. Kiev. \textbf{20}, 1--19 (1880).

\bibitem{Pinney} E.~Pinney, \newblock The nonlinear differential equation $%
y^{\prime \prime 3}=0$, \newblock Proc. Amer. Math. Soc. \textbf{1}, 681(1)
(1950).

\bibitem{Hone} A.~Hone, \newblock Exact discretization of the Ermakov-Pinney
equation, \newblock Phys. Lett. A \textbf{263}, 347--354 (1999).

\bibitem{Hawkins} R.~M. Hawkins and J.~E. Lidsey, \newblock Ermakov-Pinney
equation in scalar field cosmologies, \newblock Phys. Rev. D \textbf{66},
023523 (2002).

\bibitem{ChoiK} J.~R. Choi and B.~H. Kweon, \newblock Operator method for a
nonconservative harmonic oscillator with and  without singular perturbation, %
\newblock Int. J. Mod. Phys. \textbf{B16}, 4733--4742 (2002).

\bibitem{ChoiK2} J.~R. Choi, \newblock Exact Wave Functions of
Time-Dependent Hamiltonian Systems Involving  Quadratic, Inverse Quadratic,
and $(1/x)p+p(1/x)$ Terms, \newblock Int. J. Theor. Phys. \textbf{42},
853--861 (2003).

\bibitem{FBM} N.~Ferkous, A.~Bounames, and M.~Maamache, \newblock %
Time-dependent Schr{\"{o}}dinger equation with non-central  potentials, %
\newblock Physica Scripta \textbf{88}, 35001--35004 (2013).

\bibitem{PhysRevD.90.084005} S.~Dey and A.~Fring, \newblock Noncommutative
quantum mechanics in a time-dependent background, \newblock Phys. Rev. D 
\textbf{90}, 084005 (2014).

\bibitem{Ioffe:2002tk} M.~V. Ioffe and H.~Korsch, 
\newblock {Nonlinear supersymmetric (Darboux) covariance of the
  Ermakov-Milne-Pinney equation}, \newblock Phys.Lett. \textbf{A311},
200--205 (2003).

\bibitem{dey2015milne} S.~Dey, A.~Fring, and L.~Gouba, \newblock Milne
quantization for non-Hermitian systems, \newblock Journal of Physics A:
Mathematical and Theoretical \textbf{48}(40),  40FT01 (2015).

\bibitem{eliezer} C.~Eliezer and A.~Gray, \newblock A note on the
time-dependent harmonic oscillator, \newblock SIAM Journal on Applied
Mathematics \textbf{30}, 463--468 (1976).
\end{thebibliography}

\end{document}